\documentclass[aps,pra,twocolumn,amsmath,amssymb,showpacs,superscriptaddress]{revtex4-1}
\usepackage{color}
\usepackage{textcomp} 
\usepackage{mathrsfs,amsmath}
\usepackage{graphicx}
\usepackage{dcolumn}
\usepackage[mathlines]{lineno}
\usepackage{hyperref}
\usepackage{bm}
\usepackage{epstopdf}
\usepackage{soul}
\usepackage{epsfig}
\usepackage{bbold}

\usepackage{booktabs}

\usepackage{changes}
\usepackage{lipsum}
\setremarkmarkup{(#2)}

\usepackage{color}

\begin{document}

\title{Optimal third-harmonic generation in an optical microcavity with
	$\chi^{(2)}$ and $\chi^{(3)}$ nonlinearities}

\author{Ming Li}
\affiliation{Centre for Optical and Electromagnetic Research, State Key
	Laboratory for Modern Optical Instrumentation, College of Optical
	Science and Engineering, Zhejiang University, Hangzhou 310058, China}
\author{Chang-Ling Zou}
\affiliation{Key Laboratory of Quantum Information, CAS, University of
	Science and Technology of China, Hefei 230026, China}
\author{Chun-Hua Dong}
\affiliation{Key Laboratory of Quantum Information, CAS, University of
	Science and Technology of China, Hefei 230026, China}
\author{Dao-Xin Dai}
\email{dxdai@zju.edu.cn}
\affiliation{Centre for Optical and Electromagnetic Research, State Key
	Laboratory for Modern Optical Instrumentation, College of Optical
	Science and Engineering, Zhejiang University, Hangzhou 310058, China}

\begin{abstract}
	Third-harmonic generation can be realized via both $\chi^{(3)}$ and
	cascaded $\chi^{(2)}$ nonlinear processes in a triply-resonant microcavity.
	It is still unknown how these processes interfere with each other
	and the optimization of the conversion efficiency still remains as
	a question. In this work, the interplay between the direct third-harmonic
	generation and the cascaded process combining of the second-harmonic
	generation and the sum-frequency generation are investigated. It is
	found that the interference effect between these two processes can
	be used to improve the conversion efficiency. By optimizing the cavity
	resonance and the external coupling conditions, the saturation of
	the nonlinear conversion is mitigated and the third-harmonic conversion
	efficiency is increased. A design rule is provided for achieving efficient
	third-harmonic generation in an optical microcavity, which can be
	generalized further to the high-order harmonic generations.
\end{abstract}
\pacs{42.65.−k, 42.82.−m, 42.65.Ky}
\maketitle

\section*{Introduction}

Nonlinear photonics based on the third-order nonlinear process ($\chi^{(3)}$)
can be used for various applications in classical and quantum fields
\cite{boyd2003nonlinear,agrawal2007nonlinear,reimer2016generation,leuthold2010nonlinear}.
Generally speaking, the efficiency of the third-order nonlinear process
is very low, due to the very weak $\chi^{(3)}$ susceptibilities in
most materials. High-Q optical microcavities are often used to enhance
the nonlinear interaction \cite{vahala2003optical,strekalov2016nonlinear,Lin:17,rong2005all},
because the optical mode is confined tightly and the density of state
is greatly enhanced. During the last decades, $\chi^{(3)}$-assisted
third-harmonic generation were studied theoretically \cite{PhysRevA.79.013812}
and realized experimentally in various types of microcavities, including
microdroplets \cite{Leach:93,PhysRevLett.78.2952}, silica toroid/microsphere/microbottle
\cite{carmon2007visible,Ismaeel:12,PhysRevLett.112.093901,Asano:16},
SiN microring \cite{Levy2011}, silicon photonic-crystal micro/nano-cavities
\cite{Galli:10} and microcavities with hybrid materials \cite{Surya:18}. 

To enhance the efficiency of the third-harmonic generation (THG),
one should increase the nonlinear susceptibility of the nonlinear
processes, reduce the mode volume, optimize the modal overlap as well
as increase the quality factor of the microcavity \cite{strekalov2016nonlinear}.
For a given material system, the size and geometry of an optimal microcavity
should be designed to satisfy the phase-matching and maximize the
mode overlap factor \cite{Galli:10,Guo2016,lin2016cavity,Surya:18,kim2016direct,zeng2016imaging}.
By using the intermodal dispersion of higher-order modes for phase
matching and optimizing these parameters, third-harmonic generation
with a efficiency of $180\%\:W^{-2}$ has been observed on an integrated
photonic chip \cite{Surya:18}. On the other hand, one should notice
that the quality factor of an microcavity is usually limited due to
the fabrication imperfection and the material absorption, which builds
a barrier to increase the efficiency of THG.

Fortunately, the mechanism of cascading different nonlinear processes
provides a promising approach for engineering the nonlinear susceptibility
of the material. For example, it has been investigated that effective
$\chi^{(3)}$ nonlinear process can be constructed with low-order
$\chi^{(2)}$ process \cite{Sasagawa2009}. In a high-Q microcavity,
multiple and cascaded nonlinear processes with high efficiency can
be observed simultaneously. People have demonstrated the second-harmonic
generation (SHG), sum-frequency generation (SFG), as well as THG could
be observed simultaneously under the multiple-resonant condition in
a cavity by introducing materials with both $\chi^{(2)}$ and $\chi^{(3)}$
nonlinearity, e.g., Lithium Niobate and Aluminum Nitride (AlN). The
cascading of SHG and SFG (SHG-SFG) can also convert photons from the
fundamental to the third-harmonic frequency, which is equivalent to
the direct $\chi^{(3)}$-assisted THG (DTHG). Recently, the cascading
of $\chi^{(2)}$ processes for achieving THG was observed experimentally
in periodically-poled $\mathrm{MgO:LiNbO_{3}}$ cavity \cite{Sasagawa2009}
and $\mathrm{LiNbO_{3}}$ microdisk \cite{liu2017cascading}. The
authors demonstrated that the cascaded effect SHG-SFG in a high $\chi^{(2)}$
cavity can exceed the essential DTHG based on the $\chi^{(3)}$ nonlinearity.
Besides, people have also oberved DTHG accompanied by Raman scattering
and four-wave mixing \cite{Chen-Jinnai:16,Liang:15,wang2016frequency,Fujii:17},
which increases the potential for multiple applications. In addition
to THG, some other cascaded nonlinear processes in cavities have also
been applied for the realization of frequency combs, broadband wavelength
conversion, quantum entanglement, etc \cite{tan2011bright,ulvila2013frequency,jung2014green,wolf2017cascaded}.

Since the TH light can be generated by both processes of $\chi^{(3)}$
and cascaded $\chi^{(2)}$, it is interesting to understand how these
processes influence each other and which one is better to maximize
the THG efficiency. In our previous work, we have demonstrated that
different nonlinear processes will interfere with each other \cite{minglichi2chi3}.
Here, we study a system based on a microring resonator (MRR) in which
both the processes of direct $\chi^{(3)}$ and cascaded $\chi^{(2)}$
are involved. The behavior of these processes and the interplay between
them are investigated quantitatively. Both regimes with a low pump
power and a high pump power are investigated. It is shown that the
interference happens when the quality factor of the MRR approaches
a critical value. The interference can be utilized to improve the
conversion efficiency. In particular, when operating with a high pump
power, one should optimize the resonant frequencies and the external
coupling ratio of the MRR to compensate the saturation effect for
achieving a maximal conversion efficiency.

\begin{figure}
	\includegraphics[width=8cm]{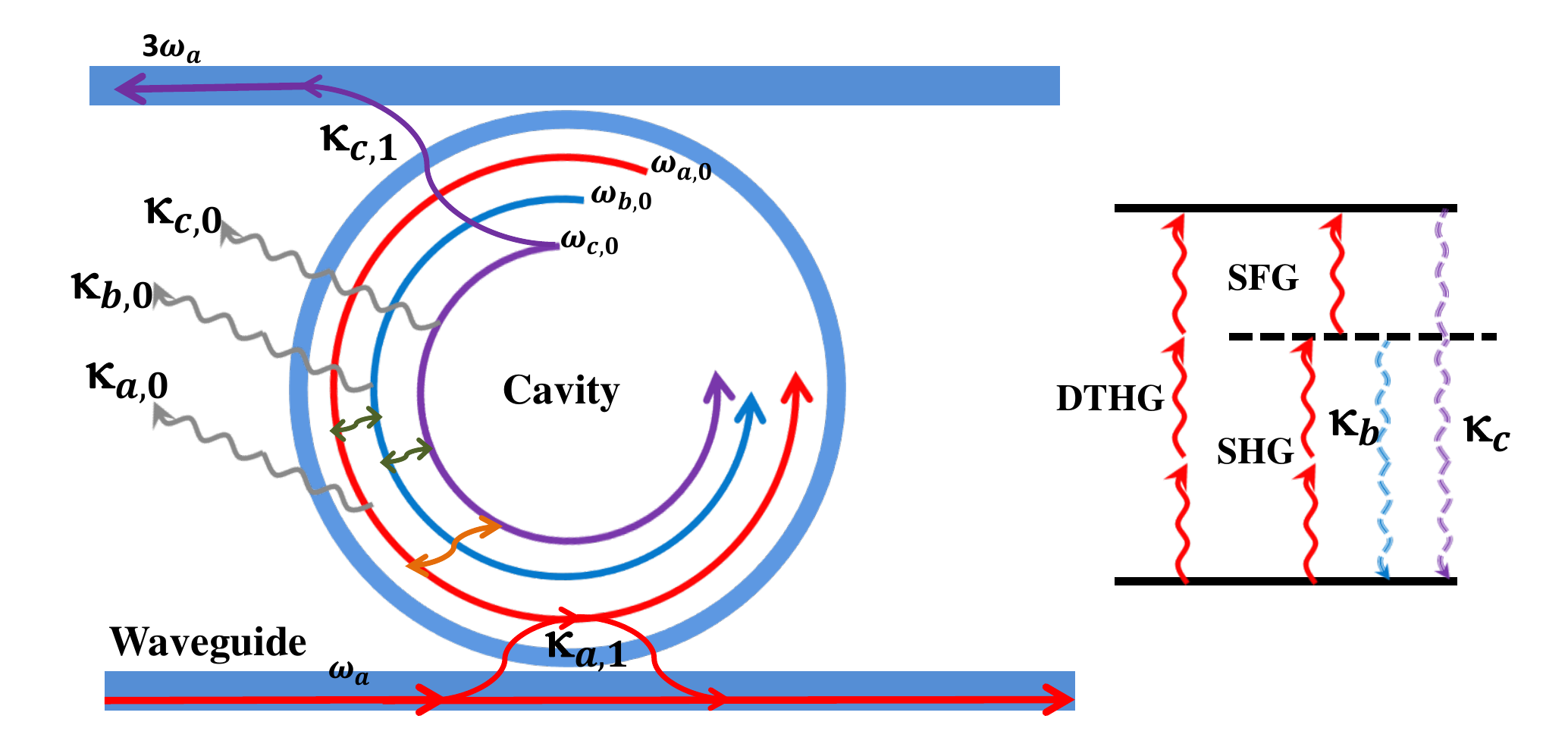}
	
	\caption{THG in a microcavity. Schematic diagram of a micro-cavity coupled
		with a waveguide. The microcavity is driven by a continuous laser
		on the fundamental mode $a$. In the microcavity, $a$ couples with
		$b$ via SHG, $a,b$ couple with $c$ via SFG, $a$ couple with $c$
		via DTHG. The DTHG and cascaded SHG-SFG can both generate photons
		in the TH mode $c$. The intrinsic and external decay rates of mode
		$i$ are $\kappa_{i,0}$ and $\kappa_{i,1}$, respectively.}
\end{figure}

\section*{System and model}

Fig.$\:$1 schematically illustrates a triply-resonant microcavity
coupled with a bus waveguide. The three resonant modes are the fundamental
(FM) mode $a$, the second-harmonic (SH) mode $b$ and the third-harmonic
(TH) mode $c$ with frequency $\omega_{i,0}$ ($i\in\{a,b,c\}$).
This cavity is formed by or filled with nonlinear materials that have
both none-zero $\chi^{(2)}$ and $\chi^{(3)}$ nonlinear susceptibilities.
The inset of Fig.$\,$1 illustrates three possible nonlinear interactions
in this cavity: (i) DTHG between $a$ and $c$ via the intrinsic $\chi^{(3)}$
nonlinearity. (ii) SHG between $a$ and $b$ via the $\chi^{(2)}$
nonlinearity. (iii) SFG between $a$, $b$ and $c$ via the $\chi^{(2)}$
nonlinearity. According to the phase-matching condition, the conservation
of angular momentum quantum number should be satisfied, i.e., $m_{a}=m_{b}/2=m_{c}/3$.
The cavity is driven by a continuous-wave laser with the frequency
$\omega_{a}$ around the resonance $\omega_{a,0}$ of the fundamental
mode. The system Hamiltonian is given by
\begin{equation}
H=H_{0}+H^{(2)}+H^{(3)}+H_{d},
\end{equation}
where $H_{0}$ is the free Hamiltonian of the three modes, $H^{(2)}$
is the nonlinear interactions due to $\chi^{(2)}$, $H^{(3)}$ is
the nonlinear interaction due to $\chi^{(3)}$, and $H_{d}$ is the
external coherent driving to the fundamental mode. One has
\begin{eqnarray}
H_{0} & = & \omega_{a,0}a^{\dagger}a+\omega_{b,0}b^{\dagger}b+\omega_{c,0}c^{\dagger}c,\\
H^{(2)} & = & g_{22}(a^{\dagger2}b+a^{2}b^{\dagger})+g_{21}(a^{\dagger}b^{\dagger}c+abc^{\dagger}),\\
H^{(3)} & = & g_{3}(a^{\dagger3}c+a^{3}c^{\dagger}),\\
H_{d} & = & \epsilon_{a}(ae^{i\omega_{a}t}+a^{\dagger}e^{-i\omega_{a}t}),
\end{eqnarray}
where $g_{3}$ is the single photon coupling strength of DTHG, $g_{22}$
is the single photon coupling rate of SHG, $g_{21}$ is the single
photon coupling rate of SFG, $\epsilon_{a}=\sqrt{2\kappa_{a,1}P_{p}/\hbar\omega_{a}}$
is the pump parameter with $\kappa_{a,1}$ and $P_{p}$ being the
external coupling rate and pump power, respectively. 

For a rotational symmetric whispering gallery microcavity, the single
photon coupling rate of the $\chi^{(2)}$ and $\chi^{(3)}$ nonlinear
processes can be calculated using the method in Ref.\cite{strekalov2016nonlinear,guo2017efficient}.
The typical values for a AlN MRR are $g_{22,21}/2\pi\sim10^{5}\,\mathrm{Hz}$,
$g_{3}/2\pi\sim10\,\mathrm{Hz}$ \cite{Guo2016,Guo2016a,guo2017parametric}.
Specially, the values can be improved greatly in a Niobate Lithium
microcavity with optimized fabrication and structures \cite{zhang2017monolithic}.
In this model, the self-phase modulation and cross-phase modulation
effects are neglected, since these effects only shift the resonance
frequencies at very high pump powers. It is convenient to solve this
problem in the rotating frame of $\omega_{a}a^{\dagger}a+2\omega_{a}b^{\dagger}b+3\omega_{a}c^{\dagger}c$
to eliminate the high frequencies, and then the free Hamiltonian
\begin{eqnarray}
H_{0} & = & \delta_{a}a^{\dagger}a+\delta_{b}b^{\dagger}b+\delta_{c}c^{\dagger}c,
\end{eqnarray}
where $\delta_{i}=\omega_{i,0}-\omega_{i}$ ($i\in\{a,b,c\}$) represents
the detuning of the resonant mode to the photon frequency. Following
the Heisenberg-Langevin equation and the mean field approximation,
both the operators and mean field obey the following form,
\begin{eqnarray}
\frac{da}{dt} & = & \alpha_{a}a-i3g_{3}a^{\dagger2}c-i2g_{22}a^{\dagger}b-ig_{21}b^{\dagger}c-i\epsilon_{a}\label{eq:a}\\
\frac{db}{dt} & = & \alpha_{b}b-ig_{22}a^{2}-ig_{21}a^{\dagger}c\label{eq:b}\\
\frac{dc}{dt} & = & \alpha_{c}c-ig_{3}a^{3}-ig_{21}ab,\label{eq:c}
\end{eqnarray}
where $\alpha_{i}=-i\delta_{i}-\kappa_{i}$, $\kappa_{i}=\kappa_{i,0}+\kappa_{i,1}$
with $\kappa_{i,0}$ and $\kappa_{i,1}$ being the intrinsic decay
rate of the cavity and the external coupling rate between the cavity
and waveguide. The intracavity photon amplitude of each mode at steady
state can be solved by setting $\frac{da}{dt}=\frac{db}{dt}=\frac{dc}{dt}=0$.
Using the input-output relationship for the microcavity $c_{out}+c_{in}=-i\sqrt{2\kappa_{c,1}}c_{s}$
\cite{walls2007quantum}, the output power of the TH light in the
waveguide and the absolute conversion efficiency can be calculated
as
\begin{eqnarray}
P_{\mathrm{TH}} & = & 2\hbar\omega_{c}\kappa_{c,1}|c_{s}|^{2},\label{eq:eff}\\
\eta & = & P_{\mathrm{TH}}/P_{\mathrm{p}},
\end{eqnarray}
respectively. According to Eqs.$\,$(\ref{eq:a}-\ref{eq:eff}), both
the internal nonlinear processes and the external coupling rates should
be optimized to achieve the maximal conversion efficiency. 

\subsection*{Weak-pump regime}

To reveal the basic physical insight of this system, we first investigate
the behavior of the system under a weak pump. In this regime, the
conversion efficiency from the fundamental mode $a$ to the harmonic
modes $b,c$ are very weak. As a result, it is resonable to apply
the non-depletion approximation on mode $a$, under which the intracavity
photon number can be treated as a constant since the backaction from
the harmonic modes are negligible. The Hamiltonian of the simplified
model reduces to
\begin{eqnarray}
H & = & \delta_{b}b^{\dagger}b+\delta_{c}c^{\dagger}c+g_{3}|a_{s}|^{3}(c+c^{\dagger})\nonumber \\
&  & +g_{22}|a_{s}|^{2}(b+b^{\dagger})+g_{21}|a_{s}|(b^{\dagger}c+bc^{\dagger}),
\end{eqnarray}
where $a_{s}=i\epsilon_{a}/\left(-i\delta_{a}-\kappa_{a}\right)$
is the intracavity photon number amplitude of the mode $a$ by neglecting
the backaction from modes $b$ and $c$. Since the phases of $a_{s}$
and the coupling strengths can be absorbed into the operators $b$
and $c$, we replace $a_{s}$ by $|a_{s}|$ in the Hamiltonian. The
dynamics of the SH and TH modes is described as, 
\begin{eqnarray}
\frac{db}{dt} & = & (-i\delta_{b}-\kappa_{b})b-ig_{22}|a_{s}|^{2}-ig_{21}|a_{s}|c\\
\frac{dc}{dt} & = & (-i\delta_{c}-\kappa_{c})c-ig_{3}|a_{s}|^{3}-ig_{21}|a_{s}|b.
\end{eqnarray}
The intracavity photon amplitude of the TH mode can be solved as
\begin{equation}
c_{s}=\frac{ig_{3}-\frac{g_{21}g_{22}}{\alpha_{b}}}{\alpha_{c}+\frac{g_{21}^{2}}{\alpha_{b}}|a_{s}|^{2}}|a_{s}|^{3}\approx(\frac{ig_{3}}{\alpha_{c}}-\frac{g_{21}g_{22}}{\alpha_{b}\alpha_{c}})|a_{s}|^{3}.\label{eq:thg}
\end{equation}
Here the first term in the bracket represents the contribution of
DTHG and the other term is due to the cascaded SHG-SFG process. It
can be seen that both the direct $\chi^{(3)}$ and the cascaded $\chi^{(2)}$
processes contribute to the generation the TH light. As an analog
to $g_{3}$, the effective third-order nonlinear coupling strength
of the cascaded SHG-SFG process can be defined as $-i\frac{g_{21}g_{22}}{\alpha_{b}}$.
Now, we can judge which process dominates the generation of the TH
light by directly comparing $g_{3}$ and $|\frac{g_{21}g_{22}}{\alpha_{b}}|$.
The coupling strength $g_{3}$ and $g_{2i}$ depend on the mode volume
and the mode overlap, which is relevant to the size and geometry of
the optical cavity. For a given optical cavity, the effective coupling
strength $|\frac{g_{21}g_{22}}{\alpha_{b}}|$ scales inversely with
the decay rate and detuning of the SH mode $b$. The higher the quality
factor of mode $b$, the stronger the effective coupling strength.
For zero detuning of mode $b$, $|\frac{g_{21}g_{22}}{\alpha_{b}}|$
reduces to $\frac{g_{21}g_{22}}{\kappa_{b}}$. The critical quality
factor $Q_{\mathrm{cr}}$, at which $g_{3}$ equals $\frac{g_{21}g_{22}}{\kappa_{b}}$,
is given as
\begin{eqnarray}
Q_{\mathrm{cr}} & = & \frac{\omega}{2}\frac{g_{3}}{g_{21}g_{22}}.
\end{eqnarray}
When $Q_{b}\gg Q_{\mathrm{cr}}$, the cascaded SHG-SFG process is
more efficient than the DTHG process. In contrast, for low quality
factor $Q_{b}\ll Q_{\mathrm{cr}}$, the DTHG process is dominant.
Therefore, it is necessary to engineer the SH mode $b$ if an high-Q
optical cavity is available. 

\begin{figure}
	\includegraphics[width=8cm]{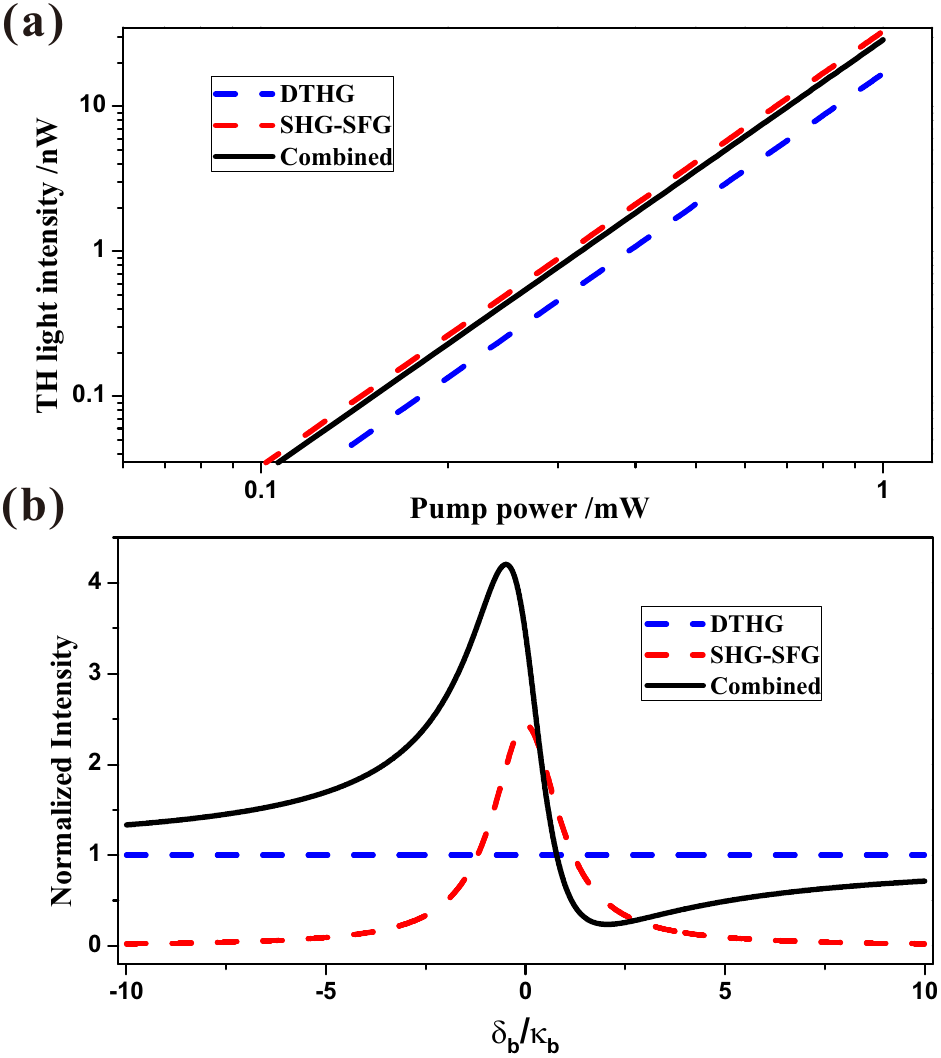}\caption{Interference between DTHG and SHG-SFG process. (a) The intensity of
		the TH mode generated from three different processes. (b) The relation
		between the normalized intensity of the TH mode and the detuning of
		the SH mode. The intensities are normalized to that of the DTHG process
		with $P_{i}^{TH}/P_{DTHG}^{TH}$ ($i$ represents the intensity of
		the TH mode from DTHG, SHG-SFG and the combined process). The parameters
		used in the calculations are $g_{3}/2\pi=10\:Hz$, $g_{2}/2\pi=0.1\:MHz$,
		$\kappa_{b}/2\pi=1\:GHz$, $\delta_{b}=0.5\kappa_{b}$, $\kappa_{a,0}/2\pi=0.4\times10^{9}\:Hz$,
		$\kappa_{b,0}=4\kappa_{a,0}$, $\kappa_{c,0}=10\kappa_{a,0}$, $\delta_{b}=0.5\kappa_{b}$.
		All the three modes are critically coupled to the waveguide.}
\end{figure}

When the quality factor is close to $Q_{\mathrm{cr}}$, the interplay
between DTHG and SHG-SFG should be taken into account. Using the nonlinear coupling strengths demonstrated
previously \cite{Guo2016,Surya:18}, the $Q_{\mathrm{cr}}$ for an
AlN MRR is calculated around $2.0\times10^{5}$ when operating at the wavelength
around $775\:nm$. This Q is achievable for the MRRs fabricated with
the current fabrication technologies. In Fig.$\,$2(a), we plot the
relationship between the intensity of the TH light and the pump power
for DTHG, SHG-SFG and the combined process with the parameters $g_{3}/2\pi=10\:\mathrm{Hz}$,
$g_{2}/2\pi=0.1\:\mathrm{MHz}$, $\kappa_{b}/2\pi=1\:\mathrm{GHz}$
and $\delta_{b}=0.5\kappa_{b}$. In this case, the efficiency of the
DTHG and SHG-SFG are at the same order of magnitude. As expected,
the efficiency of the combined process is not a direct summation of
that of the cascaded SHG-SFG and DTHG process, which demonstrates
the existence of quantum interference between nonlinear frequency
conversion processes. The interference is also shown in Eq.$\,$(\ref{eq:thg}),
in which the contribution of each process is complex. Since the photon
number in mode $c$ is proportional to $|ig_{3}-\frac{g_{21}g_{22}}{-i\delta_{b}-\kappa_{b}}|^{2}$,
the value and sign of $\delta_{b}$ determines the interference pattern
and visibility. 

The crucial factor that accounts for the interference is the phase
difference $\Delta\theta$ between $ig_{3}$ and $-\frac{g_{21}g_{22}}{\alpha_{b}}$.
For $-\pi/2<\Delta\theta<\pi/2$, DTHG and cascaded SHG-SFG process
interfere constructively and the efficiency can be enhanced. Oppositely,
the destructive interference suppresses the generation of the TH light.
We can define the effective third-order coupling strength $g_{3}^{\mathrm{eff}}=ig_{3}-\frac{g_{21}g_{22}}{-i\delta_{b}-\kappa_{b}}$
to describe the whole process. Fig.$\,$2(b) shows the relationship
between the intensity of the TH light (normalized to the DTHG process)
and the detuning of the SH mode. Extraordinarily, THG is not the most
efficient for zero detuning of mode $b$, at which the cascaded $\chi^{(2)}-\chi^{(2)}$
is the most efficient (Dashed Red line in Fig$\,$2(b)). In the case
shown in Fig.$\,$2(b), the THG efficiency can be even suppressed
up to an order of magnitude. Therefore, one must carefully control
the detuning of the intermediate mode $b$ to optimize the conversion
efficiency.

\subsection*{Strong-pump regime}

When the intracavity photon number of the pump increases, the cooperativity
of SHG $C_{\mathrm{SHG}}=\frac{g_{22}^{2}|a_{s}|^{2}}{|\alpha_{b}|^{2}}$
or DTHG $C_{\mathrm{DTHG}}=\frac{g_{3}^{2}|a_{s}|^{4}}{|\alpha_{c}|^{2}}$
approaches to unity. The nonlinear coupling induced changes of the
decay rates of the cavity modes become significant. The down-conversion
from the harmonic modes to the fundamental mode will suppress the
up-conversion. Non-depletion approximation is not valid for large
pump powers and the conversion efficiency saturates. The saturation
effect has been experimentally observed for SHG \cite{Guo2016} and
theoretically predicted for THG \cite{yu2017enhanced}. In this situation,
Eqs.$\,$(\ref{eq:a})-(\ref{eq:c}) must be solved without approximation
to analyze the behavior of the system. Then, the intracavity photon
amplitude of mode $a$ can be derived as
\begin{align}
\left[-\frac{g_{21}^{2}}{\alpha_{b}^{*}}\frac{-ig_{3}-\frac{g_{21}g_{22}}{\alpha_{b}^{*}}}{\alpha_{c}^{*}+\frac{g_{21}^{2}}{\alpha_{b}^{*}}|a|^{2}}\frac{ig_{3}-\frac{g_{21}g_{22}}{\alpha_{b}}}{\alpha_{c}+\frac{g_{21}^{2}}{\alpha_{b}}|a|^{2}}|a|^{6}+\frac{2g_{22}^{2}}{\alpha_{b}}|a|^{2}\right.\nonumber \\
\left.\left(i3g_{3}-\frac{2g_{21}g_{22}}{\alpha_{b}}+\frac{g_{21}g_{22}}{\alpha_{b}^{*}}\right)\frac{ig_{3}-\frac{g_{21}g_{22}}{\alpha_{b}}}{\alpha_{c}+\frac{g_{21}^{2}}{\alpha_{b}}|a|^{2}}|a|^{4}+\alpha_{a}\right] & =i\epsilon_{a}\label{eq:as}
\end{align}
and the amplitude of the harmonic modes are
\begin{eqnarray}
c & = & \frac{ig_{3}-\frac{g_{21}g_{22}}{\alpha_{b}}}{\alpha_{c}+\frac{g_{21}^{2}}{\alpha_{b}}|a|^{2}}a^{3}\label{eq:cs}\\
b & = & \frac{i}{\alpha_{b}}\left(g_{22}+g_{21}\frac{ig_{3}-\frac{g_{21}g_{22}}{\alpha_{b}}}{\alpha_{c}+\frac{g_{21}^{2}}{\alpha_{b}}|a|^{2}}|a|^{2}\right)a^{2}.\label{eq:bs}
\end{eqnarray}

Equation$\,$(\ref{eq:cs}) has the similar form as Eq.$\,$(\ref{eq:thg})
with an additional intensity dependent $\frac{g_{21}^{2}}{\alpha_{b}}|a|^{2}$
term in the denominator, which was neglected for weak pump powers.
Also, the intracavity photon number in the fundamental mode no longer
scales linearly with the pump power with quite a portion of energy
being converted to the harmonic modes. By numerically solving Eqs.$\,$
(\ref{eq:as})-(\ref{eq:cs}), the relationship between the conversion
efficiency and the pump power $P_{p}$ is plotted in Fig.$\,$3(a).
The blue, red and black lines represent the conversion efficiency
of the DTHG, SHG-SFG and the combined process, respectively. It can
be seen that the efficiency of THG from neither DTHG nor cascaded
SHG-SFG no longer scale cubically with the pump power when the conversion
efficiency is high. Due to the saturation, not only the SHG is suppressed,
but also the SFG efficiency from mode $a$ to $c$ decreases. Because
$g_{3}\ll g_{2}$, $C_{\mathrm{THG}}$ is much smaller than $C_{\mathrm{SHG}}$,
the saturation of the cascaded SHG-SFG appears ahead of the DTHG along
with the increase of the pump power. The efficiency of cascaded SHG-SFG
increases much slower than DTHG, which can be found by comparing the
blue and red lines. At very high pump powers, the efficiency of DTHG
can even exceed that of the SHG-SFG.

Since the saturation behaviors of the direct $\chi^{(3)}$ and the
cascaded $\chi^{(2)}$ nonlinear processes are different, their contributions
to the whole process scale differently with the pump power. The interference
will be different for high pump powers. Fig.$\,$3(b) shows the conversion
efficiency against the detuning $\delta_{b}$ of the SH mode for different
pump powers. As a consequence of the unsynchronized saturation effect
of the cascaded SHG-SFG and DTHG processes, the interference fringe
becomes less pronounced. The maximal enhancement factor decreases
as the pump increases and the optimal detuning $\delta_{b,m}$ for
maximum THG efficiency shifts to larger values when the pump increases
(Fig.$\,$3(c)). For very high values of the pump power, the individual
DTHG is efficient enough. The presence of the SH mode and their coupling
add nonlinear losses to the TH mode, thus reducing the conversion
efficiency. In this case, tuning the SH mode off the resonance is
more favorable. Simultaneously, the TH mode also add nonlinear loss
to the SH mode and the SHG efficiency is also reduced. 

To summarize briefly, the efficiency of THG is determined in two ways
in the high-pump regime. On the one hand, the generation of the TH
photons via the cascaded SHG-SFG process and the DTHG interfere with
each other. On the other hand, the cascaded SHG-SFG and DTHG process
change the decay rates of the fundamental and TH modes. The cavity
modes experience an additional nonlinear decay channel, with the total
decay rate $\kappa=\kappa_{0}+\kappa_{1}+\kappa_{nl}$, which suppresses
the nonlinear frequency conversion. Since the saturation of these
two process are unsynchronized, the constructive interference is also
destroyed.

\begin{figure}
	\includegraphics[width=8cm]{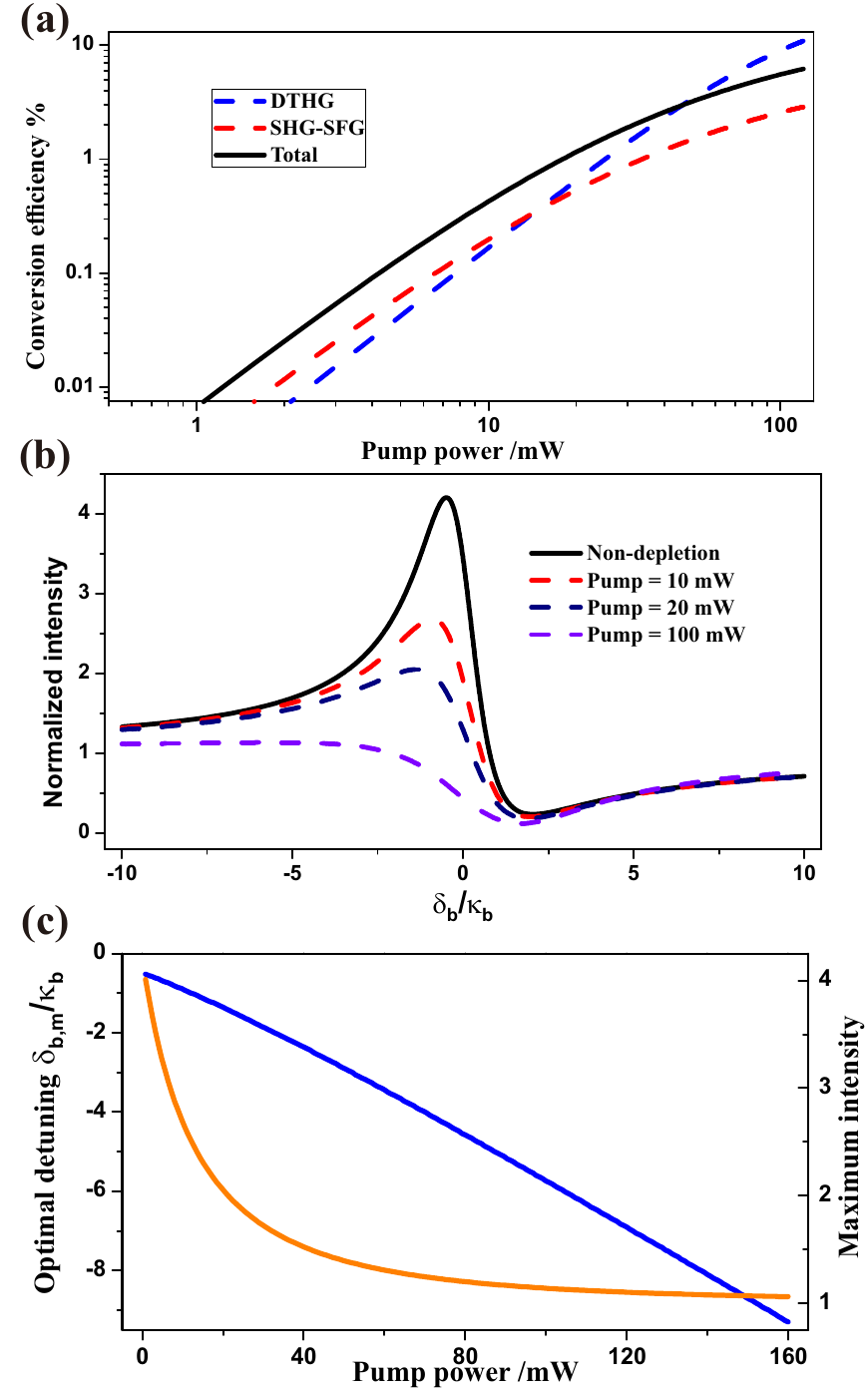}\caption{Saturation effects in TH light generation. (a) Relation between the
		conversion efficiency from the fundamental frequency to the TH frequency
		for three different processes. The cascaded SHG-SFG saturates first
		when increasing the pump power. (b) Interference fringes for different
		pump powers. The saturation of the SHG-SFG process destroys the constructive
		interference significantly. (c) Optimal detuning (Blue line) of the
		SH mode for TH light generation and the corresponding maximum enhancement
		factor (Orange line) over the DTHG process. When the pump increases,
		the optimal detuning shifts to larger values and the constructive
		interference is destroyed. In the high-pump regime, THG is dominated
		by the DTHG process.}
\end{figure}

\section*{Optimal third-harmonic generation}

In this section, we come to a more realistic question for practical
applications that what is the upper limit of the THG efficiency and
how to achieve that efficiency if the pump power and the intrinsic
quality factor of the microcavity are limited. In the above sections,
we have explicitly pointed out that the interference between DTHG
and cascaded SHG-SFG process can be used to improve the THG efficiency
with low pump powers. While operating with high pump powers, the enhancement
provided by the interference is weakened. Also, the decay rates or
quality factors of the cavity mode decreases due to the nonlinear
coupling \cite{Li:12,Guo2016}. 

For experiments, several parameters are fixed due to the limitations
of material and fabrication, therefore the nonlinear coupling strengths
$g_{21,22,3}$ and the intrinsic quality factor $Q_{i,0}=\frac{\omega_{i}}{2\kappa_{i,0}}$
of the three resonant modes ($i\:\epsilon\:\{a,b,c\}$) are limited.
So, we can engineer the mode frequency detuning $\delta_{b}$ and
the external coupling strengths $\kappa_{i,1}$ to optimize the THG.
The role of $\delta_{b}$ is to control the interference between the
DTHG and the SHG-SFG processes, while the $\kappa_{i,1}$ balances
the extraction of the TH light from the cavity $\left(\kappa_{c,1}/\kappa_{c}\right)$,
the cooperativity of the nonlinear frequency conversion and also the
saturation effect. 

Fig.$\,$4(a) shows the absolute conversion efficiency $\eta$ for
different external coupling rates with pump power $P_{\mathrm{p}}=10\:\mathrm{mW}$
(weak pump) and $P_{\mathrm{p}}=1000\:\mathrm{mW}$ (strong pump).
The detuning $\delta_{b}$ is chosen to be the optimal detuning as
studied in Fig.$\,$3(b). It is found that, the critical coupling
for both the fundamental mode and the TH mode is required to achieve
maximum conversion efficiency for weak coupling regime. The reason
is that the cooperativities $C_{THG}\ll1$ and $C_{SHG}\ll1$ in this
regime, the backaction induced decay rate $\kappa_{i,nl}$ (saturation
effect) can be neglected compared to $\kappa_{i,0}$. In this case,
the optimal external coupling rate that balances the extraction and
cooperativity is $\kappa_{c,1}\approx\kappa_{c,0}$ and $\kappa_{a,1}\approx\kappa_{a,0}$.
Since there is no input or output required for the mode $b$, thus
we want $\kappa_{b,1}\approx0$.

When operating with a high pump power $P_{\mathrm{p}}=1000\:\mathrm{mW}$,
the maximum conversion efficiency $51.7\:\%$ is obtained at $\kappa_{a,1}=2.95\:\kappa_{a,0}$
and $\kappa_{c,1}=5.01\:\kappa_{c,0}$, which lies in the over-coupled
region. If the external coupling rates are set as $\kappa_{a,1}=\kappa_{a,0}$
and $\kappa_{c,1}=\kappa_{c,0}$, the conversion efficiency is only
$25.0\:\%$. This indicates that the cooperativities approaches to
$1$, leading to relatively large changes of the effective decay rates
of the modes, $\kappa_{i,nl}\sim\kappa_{i,0}$. Therefore, the loaded
quality factors of mode $a$ and $c$ will decrease and the waveguide-cavity
coupling rates must be optimized to a higher value to ensure the critical
coupling condition $\kappa_{a,1}=\kappa_{a,0}+\kappa_{a,nl}$ and
$\kappa_{c,1}=\kappa_{c,0}+\kappa_{c,nl}$.

\begin{figure}
	\includegraphics[width=8cm]{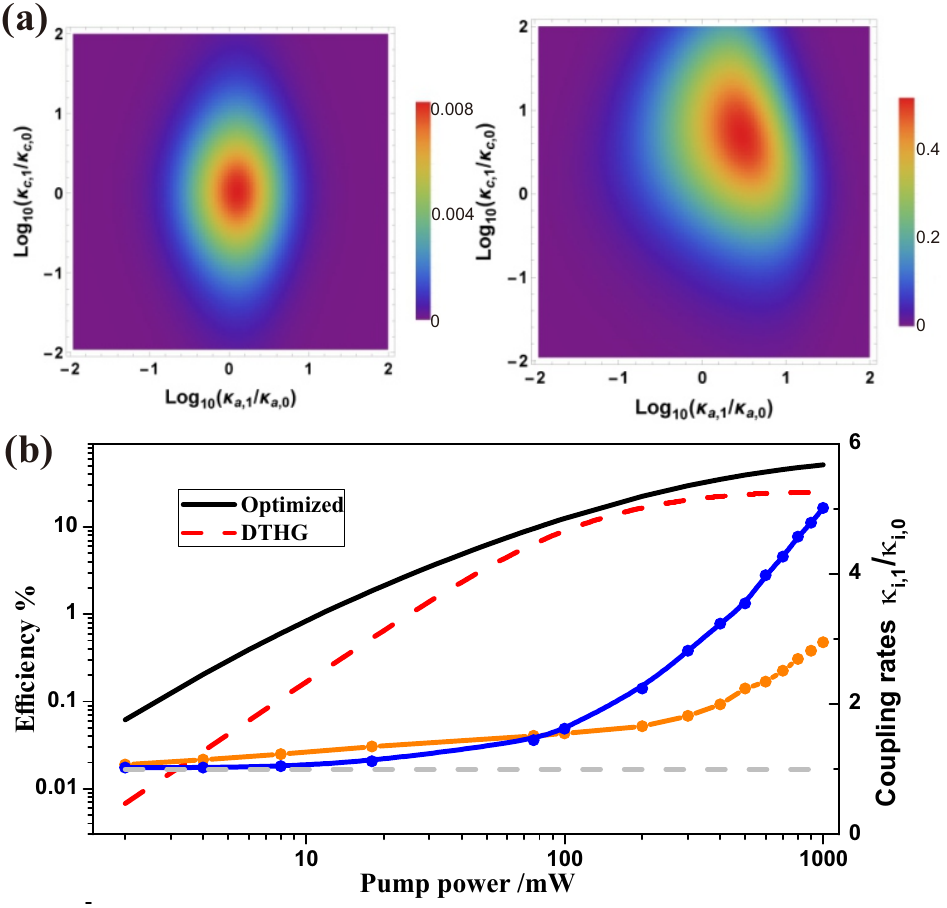}
	
	\caption{Optimization of the conversion efficiency. (a) Relationship between
		the conversion efficiency and the external coupling rates $\kappa_{a,1}$
		and $\kappa_{c,1}$. Left: $P_{\mathrm{p}}=10\:mW$. Right: $P_{\mathrm{p}}=1000\:mW$.
		In the high-pump regime, the cavity modes should be designed over-coupled
		to the waveguide to achieve optimal efficiency. (b) Relationship between
		the optimal THG efficiency and the pump power. Black line: the detuning
		$\delta_{b}$, external coupling rates $\kappa_{a,1}$ and $\kappa_{c,1}$
		are optimized; Red line: efficiency of DTHG with all the modes critical-coupled
		to the waveguide. Orange: optimal external coupling rate of the fundamental
		mode $a$. Blue: optimal external coupling rate of the TH mode $c$.
		The external coupling rates shown in this figure are normalized to
		the corresponding intrinsic decay rates $\kappa_{i,0}$.}
\end{figure}

Taking both the internal and external coupling parameters into consideration,
we calculate the achievable conversion efficiency of THG using a microcavity
with given pump power. By optimizing the detuning $\delta_{b}$ of
the SH mode $b$ and the external coupling rates of the fundamental
and TH modes, we obtain the relationship between the pump power and
the maximum conversion efficiency, as shown in Fig.$\,$4(b). The
dashed red line shows the efficiency of DTHG without optimal condition
for the external coupling. By comparing the optimal and DTHG process,
it is beneficial to utilize the cascaded SHG-SFG and the constructive
interference effect when the pump power is low. As the pump increases,
the advantage of the cascaded process and the interference is weakened
by the saturation and the two lines get close to each other. Therefore,
there is no need to design the cascaded SHG-SFG process if the pump
is too high since DTHG dominates the TH light generation. In the top
right of the Fig.$\,$4(b), the separation of the two lines becomes
large again. In this area, the higher conversion efficiency mainly
attributes from the optimized external coupling conditions. 

\section*{Conclusion}

In conclusion, we have studied the third-harmonic generation in an
optical microcavity, which allows both the $\chi^{(3)}$-assisted
direct THG and $\chi^{(2)}$-assisted cascaded SHG and SFG for the
generation of the third-harmonic light. Under the non-depletion approximation
for weak pump, we explicitly show the contributions of direct THG
and the cascading of SHG and SFG processes and predict the interference
between them. According to our analysis, the detuning of the intermediate
second-harmonic mode plays a significant role on the conversion efficiency.
For high pump powers, we analyze the different saturation effects
of these processes. The interference fringe becomes less pronounced
and the external coupling condition should be optimized to improve
the THG efficiency. At last, the optimal conversion efficiency for
a microcavity with given parameters and pump power is studied. Our
results clearly explain the mechanisms of THG in presence of both
$\chi^{(2)}$ and $\chi^{(3)}$ nonlinearity and give guidelines to
optimize the nonlinear conversion efficiency. Since our study is based
on the parameters of microcavities with current fabrication and experimental
technologies, we believe the new phenomenon predicted in our work
will be experimentally demonstrated and find their applications in
the near future. The analyses can be also generalized to other nonlinear
processes, such as SHG, Raman, and high-order harmonic generations
\cite{moore2011continuous}, which would stimulate more new physics
with cascaded nonlinear optics effects.. 
\begin{acknowledgments}
	This work was supported by National Natural Science Foundation of
	China (NSFC) (61725503, 61431166001, 11861121002, 61505195), Zhejiang
	Provincial Natural Science Foundation (Z18F050002), China Postdoctoral
	Science Foundation (No. 2017M621919).
\end{acknowledgments}

\end{document}